# A Dialogue Concerning Two World Systems:

# Info-Computational vs. Mechanistic

Gordana Dodig-Crnkovic & Vincent C. Müller
Mälardalen University & Anatolia College/ACT
www.idt.mdh.se/personal/gdc & www.typos.de

18th October, 2009

*Abstract:* The dialogue develops arguments for and against adopting a new world system – info-computationalist naturalism – that is poised to replace the traditional mechanistic world system. We try to figure out what the info-computational paradigm would mean, in particular its pancomputationalism. We make some steps towards developing the notion of computing that is necessary here, especially in relation to traditional notions. We investigate whether pancomputationalism can possibly provide the basic causal structure to the world, whether the overall research programme appears productive and whether it can revigorate computationalism in the philosophy of mind.

1.  Introduction

1.1.    Galileo, Ptolemy, Mechanicism and Systèmes du Monde

In his 1632 *Dialogue Concerning the Two Chief World Systems (Dialogo sopra i due massimi sistemi del mondo)*, Galileo contrasts two different world views: the traditional Ptolemaic geocentric system where everything in the Universe circles around the Earth, vs. the emerging Copernican system, where the Earth and other planets orbit the Sun. Even though the question whether the Earth was the center of the Universe or not was important in itself, the real scientific revolution was going on in the background; the transition from qualitative Aristotelian physics to the Galileo-Newtonian quantitative mechanistic physics necessary to support the new worldview. The new model with equations of motion for celestial bodies following Newton's laws set the standard for all of physics to come. This mechanistic paradigm accommodates even for Quantum Mechanics and Theory of Relativity, two theories that are both part of the classical mechanical "Clockwork Universe" and question its basic intuitions about a perfectly intuitive, regular and predictable "World-Machine" (Machina Mundi).

**The mechanistic world view** can is based on the following principles:

(M1) The ontologically fundamental entities of the physical reality are physical structures (*space-time & matter*) and change of physical structures *(motion)*.

(M2) All the properties of any complex physical system can be derived from the properties of its components.

(M3) Change of physical structures is governed by laws.

(M4) The observer is outside of the system observed.

Mechanistic models assume that the system is closed, isolated from the environment, and laws of conservation (energy, mass, momentum, etc.) thus hold. Environment, if modelled at all, is treated only as a perturbation for the steady state of the system.

1.2. Info-Computational Naturalism (ICON)

What we begin to see at present is a fundamentally new paradigm of not only sciences but even a more general paradigm of the universe, comparable in its radically novel approach with its historical predecessors the Mytho-poetical Universe and the Mechanistic Universe. We identify this new paradigm as Info-Computational Universe.

According to info-computational naturalism (ICON) the physical universe is fundamentally an informational structure whose dynamics are identified as computational processes (Dodig-Crnkovic 2006; 2008). This computation process is Natural computing; see Bruce MacLennan's article in this volume. Mark Burgin's article, "Information Dynamics in a Categorical Setting", presents a common framework for information and computation, building a mathematical stratum of the general theory of information based on category theory.

A remarkable feature of info-computationalism is its ability to unify living and nonliving physical world and to provide clues to mental capacities in humans and animals. Of all grand unifications or *systèmes du monde* as Greg Chaitin says in his *Epistemology as Information Theory: From Leibniz to Ω* (Chaitin 2007a) this is the first one holding promise to be able to explain and simulate in addition to non-living universe even the structure and behavior of living organisms, including the human mind.

Complexity is important for many physical phenomena, and is an essential characteristic of life, the domain in which the info-computational approach best shows its full explanatory power. Living organisms are complex, goal-oriented autonomous information-processing systems with ability of self-organization, self-reproduction (based on genetic information) and adaptation. They (we) evolved through pre-biotic and biological evolution from inanimate matter. The understanding of basic info-computational features of living beings has consequences for many fields, especially information sciences, cognitive science, psychology, neuroscience, theory of computing, artificial intelligence and robotics but also biology, sociology, economics and other fields where informational complexity is essential.

Being on the edge of a brand new era we have a good enough reason to follow Galileo's example and try to contrast two world systems – the existing and well established mechanistic framework with the new emerging unfinished but promising info-computational one.

The info-computational world view is based on the following principles:

(IC1) The ontologically fundamental entities of the physical reality are *information* (structure) and *computation* (change).

(IC2) Properties of a complex physical system cannot be derived solely from the properties of its components. Emergent properties must be taken into account.

(IC3) Change of informational structures is governed by laws.

(IC4) The observer is a part of the system observed.

Info-computational models comprise *open systems* in communication with the environment. The environment is a constitutive element for an open complex info-computational system. A *network* of interconnected parts is a typical configura-tion, where understanding is sought on the meta-level with respect to constituent parts. Info-computational models include mechanistic ones as a special case when interaction of the system with the environment may be neglected.

In what follows we will try to contrast the mechanistic and info-computational positions. This dialogue between Müller (VCM) and Dodig-Crnkovic (GDC) is the result of a series of discussions on the topic we had on different occasions over the last couple of years.

## 2. Pancomputationalist claims

### 2.1. VCM

When both authors were invited to contribute a debate to a conference in 2008[1], we jointly submitted an abstract that included the following characterization:

> Info-computationalism is the view that the physical universe can be best understood as computational processes operating on informational structure. Classical matter/energy in this model is replaced by information, while the dynamics are identified as computational processes. In this view the universe is one gigantic computer that continuously computes its next states, following physical laws.

Info-computationalism here appears as a conjunction of two theses: one about processes (computation) and one about structure (information). In this dialogue, I want to focus on the first one, that all processes are computational, which I shall call "pancomputationalism"[2]. In any case, if this pancomputationalism fails, the stronger thesis of info-computationalism fails with it.

Our first task is to gain a better understanding of the thesis involved. I shall propose some alternative readings that will be further elucidated and evaluated in this dialogue. I will start with the strongest thesis and move to weaker ones; so if one agrees with a particular thesis on this list, one will probably agree with all that follow further down.

One reading of the basic thesis in the above quote is:

P1: The universe is a computer

---

[1] "Philosophy's Relevance in Information Science" at the University of Paderborn, 3.-4.10.08, organized by Ruth Hagengruber. http://groups.uni-paderborn.de/hagengruber/pris08
[2] The term "pancomputationalism" was probably coined by (Floridi 2004: 566)

This seems to be the strongest version, so I shall call it *"strong pancomputationalism"*. Perhaps I should mention that this view normally includes the thesis that the universe is physical, something that we shall just assume in the following. A bit more restricted is the reduction to changes or 'processes' in the universe:

    P2:    All processes are computational processes

This thesis is the main target of our discussion here, so I shall just call it *"pancomputationalism"*. Very often, the point of the theory in question, however, is not what processes *are* (whatever that may mean, exactly), but how they can be *described*, so this suggests another formulation:

    P3:    All processes can be described as computational processes

*Weak pancomputationalism.* This formulation, however is ambiguous as it stands, and I shall thus avoid it. Its ambiguity stems from the fact that – as I will explain presently - there are very different reasons to claim that processes can be described in this way; reasons concerning the theory of computing and reasons concerning the nature of the universe. Whether one wants to take a realist or an anti-realist view of computing will be decisive here. Reasons concerning the nature of the universe (realist reasons) might be formulated as follows:

    P4:    All processes can be described as computational processes because we discovered that they are computational

This theory can justly bear the title of a pancomputationalism because it claims to have discovered a fundamental fact about the world. It is, in fact, just the pancomputationalism of P2 plus the claim that this feature has been discovered, so I think P4 can be disregarded; it is just 'pancomputationalism'. Another possible explanation of P3 that relies on a particular (anti-realist) notion of computation is:

    P5:    All processes can be described as computational processes because there is nothing more to being a computational process than being described as such

*Anti-realist weak pancomputationalism.* This theory does not claim that the universe has a particular structure; in fact it is often used to argue *against* a theory that a part of the world (the mind) is computational in any substantial sense. Instead, it stems from the anti-realist view that it is our description as such that makes a process into a computational one. Versions of this tradition are represented, for example, by David Chalmers (Chalmers 1993; 1994; 1996) John Searle (Searle 1980; 1990; 1992: 207f) and Oron Shagrir (Shagrir 2006). While P5 thus has a lot of sup-port in the literature, I would suggest that it is too weak for a substantial pancomputationalism in the sense envisaged by Dodig-Crnkovic.

What we need in our development of P3 is a realist formulation, like this one:

> P6: All processes can be described as computational processes because this happens to be a useful way of describing them in scientific theory

*Realist weak pancomputationalism.* This thesis takes a realist view of computation and then claims that all the actual processes of the universe are such that they can be described as computational in a scientific theory of the universe (while some processes in other possible worlds might not have this feature). It is 'weak' because it only talks about ways of description, not about realist ontology – unlike P2.

If we wanted to regard the issue just in terms of how things may be described without claiming that this description is or should be part of a scientific theory, then we would descend to what is really just a metaphorical remark, and thus the end of our sequence:

> P7: All processes can be described as if they were computational processes

*Metaphorical pancomputationalism.* This thesis is probably true and it looks like it might be extremely useful in areas as different as economics and microbiology. It does not say anything about the world, however, but that things may be described *as if* they were computational – be that scientific or not.

A further question is whether the claim of pancomputationalism, in one of the versions above, is meant to be a claim about 'everything' or 'everything deep down'; i.e. is computation a fundamental property of the universe ('deep down') and other properties relate to it systematically, e.g. they reduce to it? Or is literally everything in the universe computational? Against the latter, stronger, claim there are many areas of the universe (social, aesthetic, mental) that do not seem computational at all.[3] Are they, or are they 'deep down'?

2.2.    GDC

> P1: The universe is a computer

---

[3] Here are some examples of social, aesthetic or mental facts that do not seem computational: "The struggle over copyright in the digital age is really a power struggle", "Her hair curled beautifully", "My breath was taken away by the sight".

As all subsequent theses P2-P6 are just the weaker versions of P1 let me focus on the strongest claim, P1 in the first place. The pancomputationalist original claim is exactly P1. *Of that which is universe we say that it is a computer.* What pancomputationalists[4] actually aim for is not only giving the universe just another name ("computer") but they suggest that universe *computes*.

It is pretty obvious that universe computer is not of the same sort as my PC, as it contains stars, rocks, oceans, living organisms and all the rest (including PC's). So *we talk about a more general idea of computing and a computer*. This question of computing in real world, the nature of computing as implemented in physics, is addressed in this volume by MacLennan and Shagrir. Cooper and Sloman discuss questions of the relationship between computing, information and mind. Needless to say, in the pancomputational universe, mind is a result of natural computation that our brains supported by bodily sensors and actuators constantly perform.

In sum, I would say that all of proposed pancomputational claims P1-P6 are correct. The universe is a computer, but an unconventional (natural) one; its dynamics (temporal evolution) best can be understood as computation and it can be described as such. There is nothing more to being computational process but being used as computation.[5]

For the last claim P7, however, I propose it to be modified. Metaphor is a figure of speech while pancomputationalism concerns the physical world. I would say that pancomputationalism is a metaphor in the same sense as Niels Bohr's liquid drop model is a metaphor of atomic nucleus. In sciences we are used to talk about *models* and for a good reason. We use model as a tool to interact with the world. So if you agree to call it *model* instead of a *metaphor*, I would agree even with P7. I propose the following:

P7':    All physical processes can be *modeled* as computational processes which I recognize as pancomputationalist claim. We will have several occasions in this dialogue to return to the question of what is computing and to discuss unconventional (natural) computing that is going on in computational universe. In principle, there seems to be no ontological hindrance to our including the system or process we try to compute among the models we use. We then get what I take to be the fundamental idea of pancomputationalism: The function governing a process is calculated by the process itself[6]. The following remark by Richard Feynman lucidly explains the idea:

> It always bothers me that according to the laws as we understand them today, it takes a computing machine an infinite number of logical operations to figure out what goes on in no matter how tiny a region of space and no matter how tiny a region of time… I have often made the hypothesis that ultimately physics will not require a mathematical statement, that in the end the machinery will be revealed and the laws will turn out to be simple. (Feynman 1965: 57)[7]

---

[4] http://en.wikipedia.org/wiki/Pancomputationalism
[5] As Kaj Børge Hansen puts it, "A computation is a thought experiment. We often increase our power and ability to do thought experiments by aiding our limited memory and imagination by symbolic representations, real and virtual models, and computers." (Personal communication.)

## 3. Are there any arguments for pancomputationalism?

### 3.1. VCM

As far as I can tell the arguments in favor of pancomputationalism have been largely intuitive, indicating that this view is useful and offers an elegant all-encompassing view of the world in terms that are well understood. This feature it shares, however, with any number of all-encompassing ideologies (like pantheism or vulgar liberalism). These intuitive arguments apply to all of P1-P7 without offering any particular support for the stronger versions. What is missing is a positive argument that out of the many overall theories *this one* is true and in *one version* of it.

### 3.2. GDC

Pancomputationalism apart from being universal has nothing to do with pantheism which is not based on scientific methods. I don't think that vulgar liberalism either should be mixed in here as it is not a theory about the universe in its entirety, so I would suggest comparison with atomism as a good example of universal theory. In natural sciences the most general theories are the best ones. Universal laws are the best laws. Being universal is nothing bad, just on the contrary! It is expected of a theory of nature to be universal.

    The central question is how *epistemologically productive* this paradigm is, as it really is a research programme (on this I share the view presented by Wolfgang

---

[6] For this formulation I thank KB Hansen.
[7] Used as the motto for the 2008 Midwest NKS Conference,
http://www.cs.indiana.edu/~dgerman/2008midwestNKSconference/index.html

Hofkirchner in this volume) whose role is to mobilize researchers to work in the same direction, within the same global framework. The majority of natural sciences, formal sciences, technical sciences and engineering are already based on computational thinking, computational tools and computational modeling (Wing 2008).

Allow me to list some arguments for paradigm change, since it was said that these are missing. Following are some of the promises of info-computationalism:

- The synthesis of the (presently alarmingly disconnected) knowledge from different fields within the common info-computational framework which will enrich our understanding of the world. Present day narrow specialization into different isolated research fields has gradually led into impoverishment of the common world view.

- Integration of scientific understanding of the phenomena of life (structures, processes) with the rest of natural world helping to achieve "the unreasonable effectiveness of mathematics" such as in physics (Wigner) even for complex phenomena like biology that today lack mathematical effectiveness (Gelfand).[8]

- Understanding of the semantics of information as a part of data-information-knowledge-wisdom sequence, in which more and more complex relational structures are created by computational processing of information. An evolutionary naturalist view of semantics of information in living organisms is based on interaction (information exchange) of an organism with its environment.

- A unified picture of fundamental dual-aspect information/computation phenomenon applicable in natural sciences, information science, cognitive science, philosophy, sociology, economy and number of others.

- Relating phenomena of information and computation understood in interactive paradigm makes it possible for investigations in logical pluralism of information produced as a result of interactive computation.[9] Of special interest are open systems in communication with the environment and related logical pluralism including paraconsistent logic.

---

[8] See Chaitin, "Mathematics, Biology and Metabiology" (Foils, July 2009) http://www.umcs.maine.edu/~chaitin/jack.html

[9] This logical pluralism is closely related to phenomena of consistency and truth; see also de Vey Mestdagh & Hoepman in this volume.

- Advancement of our computing methods beyond the Turing-Church paradigm, computation in the next step of development becoming able to handle complex phenomena such as living organisms and processes of life, knowledge, social dynamics, communication and control of large interacting networks (as addressed in *organic computing* and other kinds of *unconventional computing*), etc.

- Of all manifestations of life, mind seems to be information-theoretically and philosophically the most interesting one. Info-computationalism (pancomputationalism + paninformationalism) has a potential to support (by means of models and simulations) our effort in learning about mind.[10]

## 4. What is computing? (I) The fragile unity of pancomputationalism

### 4.1. VCM

There are several theories about what constitutes a "computation", the classical one being Turing's, which identifies computation with a digital algorithmic process (or "effective procedure"). If, however, pancomputationalism requires a larger notion of computing that includes analog computing and perhaps other forms, it would seem necessary to specify what that notion is – while making sure that Turing's notion is included. It is far from clear that there is a unifying notion that can cover all that the pancomputationalist wants, and therefore there is a danger that the advertised elegance of a single all-encompassing theory dissolves under closer inspection into a sea of various related notions.

It should be noted for fairness, however that while it is clearly a desideratum to specify the central notions of one's theory, pancomputationalism can hardly be faulted for failing to achieve what is generally regarded as a highly demanding task, namely a general specification of computation.

---

[10] On the practical side, understanding and learning to simulate and control functions and structures of living organisms will bring completely new medical treatments for all sorts of diseases including mental ones which to this day are poorly understood. Understanding of information-processing features of human brain will bring new insights into such fields as education, media, entertainment, cognition etc.

One possible response to this challenge deserves a mention here, namely the response that relies on the traditional use of mathematical or formal tools in science. This could be condensed into the following thesis:

F: All physical processes can be described formally

I suspect that sympathies for this view stand behind much of the support for pancomputationalism. However, F is not identical to strong pancomputationalism and not even easy to reconcile with it, for three reasons: a) F talks about the possibility to *describe* things, i.e. it does not make any claim to a realist reading [unlike P1 and P2], b) its use would identify computing with formal description and c) it explicitly talks about processes and thus forbids any swift moves from P2 to P1 – in case that P1 is the desired thesis. What is the logical relation between pancomputationalism and thesis F?

4.2.     GDC

Actually the lack of understanding for what computing is may be a good argument for starting this whole research programme. At the moment, the closest to common acceptance is the view of computing as information processing, found in Neuroscience, Cognitive science and number of mathematical accounts of computing; see (Burgin 2005) for exposition. For a process to be a computation a model must exist such as algorithm, network topology, physical process or in general any mechanism which ensures predictability of its behavior.

The three-dimensional characterization of computing can be made by classification into orthogonal types: digital/analog, interactive/batch and sequential/parallel computation. Nowadays digital computers are used to simulate all sorts of natural processes, including those that in physics are described as continuous. In this case, it is important to distinguish between the mechanism of computation and the simulation model. It is interesting to see how computing is addressed in the present volume, especially Barry Coopers account of definability and Bruce MacLennans embodied computing. We will mention symbolic vs. subsymbolic computing as important in this context.  So symbolic part is what is easily recognized as thesis F[11]. In a sense we may say that F applies even to sub-

---

[11] As already pointed out, we have two different types of computation: physical substrate subsymbolic and symbolic which also is based on physical computation. In Mark Burgin's words: "However, we can see that large physical computations can give a futile symbolic result, while extended and sophisticated symbolic computations sometimes result in meager physical changes. Pancomputationalism actually cannot exist without accepting the concept of physical computation. Discovery of

symbolic computing on a meta-level. What is described formally is not the computational process step by step, but the mechanism that will produce that process.

*Information processing*[12] is the most general characterization of computing and common understanding of computing in several fields. In the info-computational approach information is a structure and computation is a process of change of that structure (Dodig Crnkovic, 2006). I have used the expression "dynamics of information" for computation. No matter if your data form any symbols; computation is a process of change of the data/structure. On a fundamental quantum-mechanical level, the universe performs computation on its own (Lloyd, 2006). Symbols appear on a much higher level of organization, and always in relation with living organisms. Symbols represent something for a living organism, have a function as carriers of meaning.[13] (See Christophe Menant in this volume).

## 5. What is computing? (II) Discrete vs. continuum or digital vs. analog

### 5.1. VCM

I used to believe that what a computational process is was nicely defined by Church and Turing in the 1930ies, namely that these are the "effective" procedures", just the algorithms that can be computed by some Turing machine. This does, at least, provide something like a 'core' notion. It can be expanded in several ways, but any notion of computing should include this 'core'. One expansion is 'hypercomputing'; the idea that there can be algorithmic procedures that compute what no Turing machine could compute (typically by carrying out infinitely many computing steps). Now, I do not think that a machine can carry out infinitely many steps in finite time and come up with an output (Müller 2008a) – but I would grant, of course, that hypercomputing is computing, if only it were physically possible in this world, or indeed in any possible world.

There might be a set of computing procedures that is larger than the one defined by Church-Turing – and there is certainly a mathematical set of computable functions larger than that computable by Turing machine (e.g. that computable by Turing's idea of his machine plus "oracle"). This is still quite far from saying that the universe is a computer (P1 above), however. So pancomputationalism

---

quantum and molecular computing shows that the same symbolic computation may result from different kinds of physical computation." (From e-mail exchange with Mark Burgin, 26.07.2009 ) probably has to add 'analog computers' as well, machines who's processing is not digital steps and who's output thus requires measurement to a degree of accuracy (if there is any 'output' at all).[14] My understanding of 'computer', as suggested by (Turing 1936), is that such machines characteristically go beyond mere *calculators* (like those already invented by Leibniz and Pascal) in that they are *universal*; they can, in principle, compute *any* algorithm, because they are programmable – in this sense, Zuse's Z3 was the first computer (1941). If this feature of universality is a criterion for being a computer, then analog machines do not qualify because they can only be programmed in a very limited sense. This is a question of conventional terminology, however, so if we want to call such analog devices 'computers', we can. What is not clear to me is how this relates to the notion of 'symbol', traditionally a central one for computing. Presumably, analog computers do not use symbols, or digital states that are interpreted as symbols.

[12] A popular account in Wikipedia http://en.wikipedia.org/wiki/Computation

[13] Douglas Hofstadter has already addressed the question of a symbol formed by other symbols in his *Gödel, Escher, Bach*. (Taddeo and Floridi 2005) present a critical review of the symbol grounding problem with a suggestion that symbols must be anchored in sub-symbolic level.

So, if we grant that computing includes digital hypercomputing and analog computing, this raises two questions: First, how can you guarantee that the notion of 'computing' you are using here is in any sense unified, i.e. *one* notion? (The question raised above.) And second, how does this broadening of the notion help to support the notion that everything is computing, beyond providing an ad hoc answer to the obvious challenge that not everything is a digital computer?

5.2.  GDC

In the above, you identify hypercomputing as one way to expand the notion of computing, by carrying on infinitely many computational steps in a finite time, so we can focus our discussion for a while on the analysis of that statement. We can translate this question in its turn into Chaitin's question about the existence of real numbers; see Chaitin *How real are real numbers?* (in Chaitin 2007b: 276). For Chaitin real numbers are chimeras of our own minds, they just simply do not exist! He is in good company. Georg Leopold Kronecker's view was that, while everything else was made by man, the natural numbers were given by God. The logicists believed that the natural numbers were sufficient for deriving all of mathematics. In the above, you seem to suggest this view.

Even though pragmatic minded people would say that discrete set can always be made dense enough to mimic continuum for all practical purposes, I think on purely principal grounds that one cannot dispense with only one part in a dyadic pair and that continuum and discrete are mutually defining.[15]

---

[14] This extension to 'analog computers' is not necessary if pancomputationalism adds the thesis of 'digital physics', that the world is fundamentally digital; something that I find rather implausible - though there are arguments about this issue (Müller 2008b) (Floridi 2009).

Here I would just like to point out that the discrete – continuum problem lies in the underpinning of calculus and Bishop George Berkeley in his book *The analyst: or a discourse addressed to an infidel mathematician*, argued that, although calculus led to correct results, its foundations were logically problematic. Of derivatives (which Newton called fluxions) Berkley wrote:

> And what are these fluxions? The velocities of evanescent increments. And what are these same evanescent increments? They are neither finite quantities, nor quantities infinitely small, nor yet nothing. May we not call them ghosts of departed quantities?[16]

Philosophical problems closely attached to the idea of infinity in mathematics are classical ones.

From physics on the other hand, there are persistent voices, such as (Lesne 2007) witnessing for the necessity of continuum in physical modeling of the world. Here is the summary:

> This paper presents a sample of the deep and multiple interplay between discrete and continuous behaviours and the corresponding modellings in physics. The aim of this overview is to show that discrete and continuous features coexist in any natural phenomenon, depending on the scales of observation. Accordingly, different models, either discrete or continuous in time, space, phase space or conjugate space can be considered. (Lesne 2007)

(Floridi 2009) proposes the Alexandrian solution to the above Gordian knot by cutting apart information from computation, and expressing everything in terms of information. This would be analog to describing a verb with a noun; it is possible but some information gets lost. It is nevertheless true that informational struc-ture of the universe is richer than what Turing Machines as a typical mechanical/mechanistic model can produce.[17]

---

[15] I suppose that this dyadic function comes from our cognitive apparatus which makes the difference in perception of discrete and continuous. It is indirectly given by the world, in a sense that we as a species being alive in the world have developed those dyadic/binary systems for discrete (number) and continuous (magnitude) phenomena as the most effective way to relate to that physical world.
Much of our cognitive capacities seem to have developed based on vision, which has on its elementary level the difference between: signal/no signal.

[16] http://www.maths.tcd.ie/pub/HistMath/People/Berkeley/Analyst/Analyst.html Berkeley talks about *the relationship between the model and the world*, not about the inner structure of the model itself. Worth noticing is KB Hansen's remark that "problems observed by Berkeley have been solved by Bolzano, Cauchy, Riemann, Weierstrass, and Robinson. Modern mathematical analysis rests on solid foundations."

> … digital ontology (the ultimate nature of reality is digital, and the universe is a computational system equivalent to a Turing Machine) should be carefully distinguished from informational ontology (the ultimate nature of reality is structural), in order to abandon the former and retain only the latter as a promising line of research. Digital vs. analogue is a Boolean dichotomy typical of our computational paradigm, but digital and analogue are only "modes of presentation" of Being (to paraphrase Kant), that is, ways in which reality is experienced or conceptualised by an epistemic agent at a given level of abstraction. A preferable alternative is provided by an informational approach to structural realism, according to which knowledge of the world is knowledge of its structures. The most reasonable ontological commitment turns out to be in favour of an interpretation of reality as the totality of structures dynamically interacting with each other. (Floridi 2009: 151)

What info-computationalist naturalism wants is to understand that dynamical interaction of informational structures as a computational process. It includes digital and analogue, continuous and discrete as phenomena existing in physical world on different levels of description and digital computing is a subset of a more general natural computing.

The question of continuum vs. discrete nature of the world is ages old and it is not limited to the existing technology. Digital philosophy as well as Turing machine has been epistemologically remarkably productive (see Stephen Wolframs work, e.g. (Wolfram 2002) along with Ed Fredkin and number of people who focused on the digital aspects of the world). Digital is undoubtedly one of the levels we can use for the description, but from physics it seems to be necessary to be able to handle continuum too (as we do in Quantum Mechanics). For a very good account, see (Lloyd 2006).

---

[17] For we talk about computational processes that not only calculate functions but are able to interact with the world, posses context-awareness, ability of self-organization, self-optimization and similar.

## 6. What is computing? (III) Natural computing as a generalization of the traditional notion of computing

### 6.1. GDC

We have already discussed *hypercomputing* as the possibility of carrying on infinitely many (computational) steps in a finite time as a question of our understanding of the nature of the world (continuous, discrete) and our idea of infinity. There is however yet another possibility to approach the question of computing beyond the Turing model which goes under different names and has different content: *natural computing, unconventional computing, analog computing, organic computing, sub-symbolic computing*, etc.

In order to expound the present understanding of computing, and its possible paths of development we study the development of the computing field in the past half a century, driven by the process of miniaturization with dramatically increased performance, efficiency and ubiquity of computing devices. However, this approach based on the understanding of computation as symbol manipulation performed by a Turing Machine is rapidly approaching its physical and conceptual limits.

Ever since Turing proposed his machine model identifying computation with the execution of an algorithm, there have been questions about how widely the Turing Machine model is applicable. Church-Turing Thesis establishes the equivalence between a Turing Machine and an algorithm, interpreted as to imply that *all of computation must be algorithmic*. Hector Zenil and Jean-Paul Delahaye in this volume investigate the question of the evidence of the algorithmic computational nature of the universe.

With the advent of computer networks, the model of a computer in isolation, represented by a Turing Machine, has become insufficient; for an overview see (Dodig Crnkovic 2006). Today's software-intensive and intelligent computer systems have become huge, consisting of massive numbers of autonomous and parallel elements across multiple scales. At the nano-scale they approach programmable matter; at the macro scale, multitude of cores compute in clusters, grids or clouds, while at the planetary scale, sensor networks connect environmental and satellite data. The common for these modern computing systems is that they are *ensemble-like* (as they form one whole in which the parts act in concert to achieve a common goal like an organism is an ensemble of its cells) and *physical* (as ensembles act in the physical world and interact with their environment through sensors and actuators).

A promising new approach to the complex world of modern autonomous, intelligent, adaptive, networked computing has successively emerged. Natural computing is a new paradigm of computing (MacLennan, Rozenberg, Calude, Bäck, Bath, Müller-Schloer, de Castro, Paun) which deals with computability in the physical world such as biological computing/organic computing, computing on continuous data, quantum computing, swarm intelligence, the immune systems, and membrane computing, which has brought a fundamentally new understanding of computation.

Natural computing has different criteria for success of a computation. The halting problem is not a central issue,[18] but instead the adequacy of the computational response. Organic computing system e. g. adapts dynamically to the current conditions of its environment by self-organization, self-configuration, self-optimization, self-healing, self-protection and context-awareness. In many areas, we have to computationally model emergence not being algorithmic (Aaron Sloman, Barry Cooper) which makes it interesting to investigate computational characteristics of non-algorithmic natural computation (sub-symbolic, analog). Interesting to observe is epistemic productiveness of natural computing as it leads to a significantly bidirectional research (Rozen); as natural sciences are rapidly absorbing ideas of information processing, field of computing concurrently assimilates ideas from natural sciences.

### 6.2. VCM

P≠NP. Or, to be a bit more explicit: I really suspect that Turing was right about his set of digitally computable functions, no matter how long it might take to compute them. All of the fashionable 'beyond Turing' computing (small, networked, natural, adaptive, etc. etc.) is either just doing what a Turing machine does or it is not digital computing at all. If it is not digital computing, then my question (notorious by now) is: Why call it computing? In what sense of that word?

### 6.3. GDC

Let me remind that for a process to be a computation a model must exist such as algorithm, network topology, physical process or *in general any mechanism which ensures predictability of its behavior*. So we distinguish computation models and physical implementations of computation.

---

[18] In the Turing model a computation must halt when execution of an algorithm has finished.

Talking about *models of computation* beyond Turing model, super-recursive algorithms are an instructive example. They represent computation which can give a result after a finite number of steps, does not use infinite objects, such as real numbers, and nevertheless is more powerful than any Turing machine. *Inductive Turing machines* described in (Burgin 2005) have all these properties. Besides, their mode of computation is a kind of a natural computation, as demonstrated with respect to evolutionary computations.

When it comes to *physical implementations*, natural computing presents the best example of the more general computational process than that used in our present days computers. In what sense of the word is that computing? In the sense of computation as a physical process, see Feynman's remark about physical computing from 2.2 above.

Physical processes can be used for digital and analog computation. It is true that historical attempts to build analog computers did not continue because of the problem with noise. In a new generation of natural computers we will use features organic computing possess in order to control complexity. Organic systems are very good at discerning information from noise.

This leads us to the next important characteristics of natural computers. They will not be searching for a perfect (context free) solution, but for a good enough (context dependent) one. This will also imply that not all computational mechanisms will be equivalent, but we will have classes of equivalence of computational devices in the same sense as we have different types of computational processes going on at different levels of organization.

Why call it computation?

Simply because it is a generalization of present day computation from discrete symbol manipulation to any sort of (discrete, continuous) manipulation of symbols or physical objects (discrete, continuous), which follow physical or logical laws.

## 7. Computation and causation

### 7.1. VCM

As we already saw, pancomputationalism seems to rely on an understanding of computation that is rather unconventional. Conventional understanding investigates a physical process and then says about that process that it computes (a function). *Which* processes in the world are the ones that are computing is a thorny question that hinges on the criteria; on whether one regards computing as a mat-ter of discovery or a matter of perception; etc. So, even if one says things like that the universe is information processing (Wiener 1961: 132 etc.), this is still meant in the sense that there is some 'stuff' in the world that is undergoing processes which are information processing – not that the universe *is* a computer.

No matter which processes are regarded as computational ones (i.e. how narrow or wide the notion of computing is taken to be), a usual assumption is that the same computation can be carried out by different physical processes – one example of this is the remark that the same software can 'run' on different hardware, even on hardware that is structurally quite different. What this underlines is that the output of a computation, e.g. "0", is a different entity from the outcome of the physical process, e.g. a switch being in "position A" (which stands for "0"). The computation is not the cause of the position of the switch, but the physical process is. The same computational process on different hardware would have resulted in "0", but quite possibly *not* in a switch in "position A". In fact, a computation cannot cause anything, it is just a syntactic event, or perhaps the syntactic description of an event (out of the massive literature on this issue, see (Piccinini 2008)). If this is right, then computation cannot be used as an overall empirical theory, as I indicated in my paper for the Paderborn meeting (Müller 2009).

### 7.2. GDC

Pancomputationalism indeed often (but not necessarily) relies on an understanding of computation that is unconventional. Exactly that unconventional computation is one of the most exciting innovations that pancomputationalism supports. Not all adherents of unconventional computation (computation beyond Turing limit) are pancomputationalists. Unconventional computation will be found all over this volume (Cooper, MacLennan, Shagrir, …) based on different arguments and approaches. There are conferences and journals on unconventional computing, organic computing, and natural computing. I see it as a good sign of coherence coming from different, often completely unrelated fields. A good overview on non-classical computation may be found in (Stepney 2005).

When it comes to the issue of causal inefficacy of computation, that is really not a problem for control systems or robotics where you indeed see computation causing an artifact to interact with the environment. Info-computationalism has no problem with computation not being causally connected with the physical world. As the world computes its own next state, it means that computation has causal power. Not only spontaneous computation of the universe in form of natu-ral computation is causally effective, even human-designed (constructed) devices controlled by computational processes show that computation is what directly connects to the world.

In the same way as there is no information without (physical) representation (Karnani et al. 2009), there is no computation without information (which must have physical representation). So any output of a computation performed by a computer (say "0" from your example) can in principle be used as an input for a control system that launches a rocket or starts any sort of machine controlled by a computer. Today we have numerous examples of embedded computers and even embodied ones (see MacLennan in this volume) where computational processes control or in other ways impact physical world.

**8.    Is pancomputationalism vacuous or epistemically productive?**

8.1.    VCM

Presumably, pancomputationalism is an empirical theory, so it should indicate which empirical evidence it will count as supportive and make predictions about empirical findings that – if they do not materialize – would count as evidence *against* the theory, perhaps even as falsification. The absence of such links to empirical findings would increase suspicion that the theory is actually devoid of content. Has the theory produced any new testable hypotheses?

8.2.    GDC

> "Theories are nets: only he who casts will catch." Novalis

Novalis is quoted by Karl Popper in the introduction to *The Logic of Scientific Discovery*. In the third chapter, Popper elaborates:

> Theories are nets cast to catch what we call "the world'': to rationalize, to explain, and to master it. We endeavor to make the mesh ever finer and finer.[19]
> (Popper 1959)

Not only so that no theory, however general, can capture all aspects of the universe simultaneously (and thus we have a multitude of different general scientific theories valid in their specific domains, on specific level of abstraction), but even more importantly: pancomputationalism is not a single monolithic theory but *a*

---

[19] This example can be paraphrased to say that not only that our nets are getting finer, but maybe altogether different methods of fishing and not only the finer-grained mesh nets can be devised.

*research programme*. We talk about *système du monde*. This volume provides examples and shows how things happen to develop more in the spirit of *Let a Thousand Flowers Bloom*. The process of consolidation, purification, formalization is the next step. We are still in a discovery phase.

In order to understand the development of a *research programme* let us return to the analogy of pancomputationalism with atomism[20], the belief that all physical objects consist of *atoms* and *void* (Leucippus and Democritus). We can equally ask how atomism could have possibly been falsified. I don't think it could. Because atomism (and in a similar way pancomputationalism) is not to be understood as a single hypothesis but as a research programme. In the strict sense atomism has already been falsified because atoms are not indivisible (but made of nucleus made of nucleons made of quarks made of …) and void (vacuum) is not empty (but full of virtual particles that pop up into being and disappear again). The process of the development of atomist research project has both changed the original idea about atoms and void and what we identify as their counterparts in the physical world.

The fundamental question thus does not seem to be about the "truth" of a singular statement such as: Is it true that there are only atoms and void? But as we use atoms and void as a net to catch the structure of the physical reality, those ideas are instrumental to our understanding, and in the interaction with the world, both our concepts and what we are able to reach to in the world change concurrently. What is fundamental is *construction of meaning*, or *epistemological productiveness* of a paradigm, or how much we can learn from the research programme.

You (Müller 2008c: 38) rightly use Kant to suggest the way to address the question of how to define *Computing and Philosophy*, namely by answering the following questions:

What can we hope for (from Computing and Philosophy)?
What should we do (with Computing and Philosophy)?
What can we know (about Computing and Philosophy)?

Equivalent questions can be asked about info-computationalist programme. A theory (or a paradigm) is an epistemic tool, that very tool Novalis and Popper use to catch (or extract as Cooper in this volume says) what for us is of interest in the world. Compared to mytho-poetic and mechanistic frameworks the emerging info-computational paradigm is the most general one and the richest in expressive repertoire developed through our interaction with the world.[21] When the dominating interaction with the world was mechanistic, the most general paradigm was mechanistic. The world in itself/for itself is simply a reservoir/resource (Floridi 2008) of possible interactions for a human. We know as much of the world as we explore and "digest" (as a species or as a community of praxis).

---

[20] I have used this analogy with atomism for many years, only recently to see in a Documentary/Drama "Victim of the brain" (on Hofstadter/Dennett's "The Mind's I" (Hofstadter and Dennett 1981) featuring Daniel Dennett and Marvin Minsky), that Douglas Hofstadter uses exactly that argument in a very elegant way, see http://www.mathrix.org/liquid/#/archives/victim-of-the-brain

> Since we wish to devise an intelligible conceptual environment for ourselves, we do so not by trying to picture or photocopy whatever is in the room (mimetic epistemology), but by interacting with it as a resource for our semantic tasks, interrogating it through experience, tests and experiments. Reality in itself is not a source but a resource for knowledge. Structural objects (clusters of data as relational entities) work epistemologically like constraining affordances: they allow or invite certain constructs (they are affordances for the information system that elaborates them) and resist or impede some others (they are constraints for the same system), depending on the interaction with, and the nature of, the information system that processes them. They are exploitable by a theory, at a given Level of Abstraction, as input of adequate queries to produce information (the model) as output. (Floridi 2008)

All we have are constructs made for a purpose, and so is even the case with pancomputationalism: let's say world is a computer, what sort of computing is it then? It is not a vacuous tautology but a proposal for exploration, a research programme. It presupposes a dynamical reflexive relationship between our understanding of the physical world and our theoretical understanding of computation or what a computer might be.

The worst thing which can happen is that some of the world is impossible to use for learning of any new principles or building any new smart machines. That may happen if physical processes are irreducible and if we want to know the result of computation we have to use the replica of a system, which is not very useful. But that is not the major issue. First of all even in case or randomness (Chaitin 2007a) when no information compression is possible the physical world shows remarkable stability and we can expect it to repeat the same behavior under same circumstances, so we don't have to actually repeat all computations, but remember recurring behaviors.

---

[21] Our nets are global computer networks of connected computational, information processing devices. The classic era of mechanism was focused on matter and energy. Our own info-computational paradigm focuses on information and computation.

## 9. Pancomputationalism and the mind

### 9.1. VCM

The view that the human mind is a computer has been a cornerstone of the cognitive sciences from their beginning, supported by the philosophical position of 'machine functionalism'. It has come under increasing pressure in recent years, and under the impression of the main arguments many have been lead to abandon it.[22] Is there any substantial sense in which info-computationalism relieves this pressure and blows some life into the notion that the mind is a computer beyond saying that everything is?

### 9.2. GDC

Yes, I would say so. I would like to claim that info-computationalism (info-computationalist naturalism) has something essentially new to offer and that is *natural computation/organic computation*, which applies to our brains too.

The classical critique of old computationalism based on abstract, syntactic notion of computation represented by Turing Machine model does not apply to the dynamic embodied physical view of computing that new natural computational models support. (Scheutz 2002) has the right diagnosis:

> Instead of abandoning computationalism altogether, however, some researchers are reconsidering it, recognizing that real-world computers, like minds, must deal with issues of embodiment, interaction, physical implementation, and semantics.

Scheutz similarly to Shagrir in this volume concludes that according to all we know brain computes, but the computation performed is not of in the first place a Turing Machine type. Several papers in this issue contribute to the elucidation of earlier misunderstandings; from Marvin Minsky's analysis of the hard problem of consciousness to Aaron Sloman's approach to mind as virtual machine. This book presents an effort to build the grounds for understanding of computing in its most general form and to use it in addressing real world phenomena, including life and mind, those topics mechanistic models are not suitable to deal with.

---

[22] A quick indication of some main points: The problem of meaning in a computational system (Chinese room and symbol grounding), the critique of encodingism (Bickhard), the stress on non-symbolic or sub-symbolic cognition, the integration of cognition with emotion and volition, the move away from a centralized notion of cognition and towards 'embodiment', etc.

Part of our previous discussion about discrete vs. continuum is relevant for the argument about computational nature of mind. If computation is allowed to be continuous, then the mind can be computational:

> Brains and computers are both dynamical systems that manipulate symbols, but they differ fundamentally in their architectures and operations. Human brains do mathematics; computers do not. Computers manipulate symbols that humans put into them without grounding them in what they represent. Human brains intentionally direct the body to make symbols, and they use the symbols to represent internal states. The symbols are outside the brain. Inside the brains, the construction is effected by spatiotemporal patterns of neural activity that are operators, not symbols. The operations include formation of sequences of neural activity patterns that we observe by their electrical signs. The process is by neurodynamics, not by logical rule-driven symbol manipulation. The aim of simulating human natural computing should be to simulate the operators. In its simplest form natural computing serves for communication of meaning. Neural operators implement non-symbolic communication of internal states by all mammals, including humans, through intentional actions. (…) I propose that symbol-making operators evolved from neural mechanisms of intentional action by modification of non-symbolic operators. (Freeman 2009)

The above shows nicely the relationship between symbolic and non-symbolic computing. All that happens inside our heads is non-symbolic computing. (Freeman claims it is non-symbolic while Shagrir with neuroscientists claims it is computing.) Our brains use non-symbolic computing internally to manipulate relevant external symbols!

If we learn to interpret life as a network of information processing structures and if we learn how our brains (and bodies) perform all that information processing then we will be able to make new computers which will smoothly connect to our information processing cognitive apparatus.

To summarize, we can choose digital description but then we will be able to see the world in that "digital light". If we choose continuum, we will capture different phenomena. Pancomputationalism does not exclude any of the (discrete, continuum, digital or analog) computing (information processing). Info-computational naturalism, being a general unifying approach connects natural information processes with corresponding informational structures.

## 10. Concluding remarks

10.1. VCM

What I think this exchange shows is that a lot of work remains to be done before we can say that pancomputationalism is a well-understood and evaluated position (not to mention info-computationalism, which involves further claims). I am therefore not of the view that the position is refuted, but that we need to clarify its claims, its fruitfulness and its possible problems – it is to this program that we hoped that our discussion would contribute (and to my mind it did).

It is in this intention that I suggested a list of possible theses at the outset. It might be useful to list them here again (where P4-P6 are possible readings of P3):

P1: The universe is a computer *(strong pancomputationalism)*

P2: All processes are computational processes *(pancomputationalism)*

P3: All processes can be described as computational processes *(weak pancomputationalism)*

P4: All processes can be described as computational processes because we discovered that they are computational (= P2)

P5: All processes can be described as computational processes because there is nothing more to being a computational process than being described as such *(anti-realist weak pancomputationalism)*

P6: All processes can be described as computational processes because this happens to be a useful way of describing them in scientific theory *(realist weak pancomputationalism)*

P7: All processes can be described as if they were computational processes *(metaphorical pancomputationalism)*

In response, we were told in no uncertain terms that, out of the various theses, strong pancomputationalism (P1) is the intended reading. Fine, in this strong realist reading the answer to my first question becomes even more urgent: What *would be* the case if the theory were false, i.e. what *would* a counterexample look like? My suggestion (in good Popperian tradition, since his name was invoked), is that there is a danger for very general ideologies that seem to explain everything, but really are empty and explain nothing. If classical atomism is still a useful or true theory (unlike pantheism), there must be a sense in which it can be interpreted as such.

In the defense of the theory, it was stressed that pancomputationalism should be viewed as a 'research program', a 'paradigm' that it is 'epistemologically productive', and that in any case theories should not be viewed as statements but as nets. All of this looks like P6, rather than P1. As long as it is granted that these two are different theses, this strategy might be accused of claiming the stronger thesis but defending the weaker one. I see several instances of this problem here.

One example is the response to the problem of the apparent causal inefficacy of computing, countered by examples of computers that do things. As an indication to show how this is not the same, let us look at Sloman's suggestion that there are 'virtual machines' active in the human brain that causally generate aspects of conscious experience (Sloman 2009). This looks like he is saying that computing has causal power – but not quite, since he says that virtual machines are mathematical objects that do nothing, only *running* virtual machines have causal powers (since they run on physical hardware, I would add). He makes the crucial distinction. My problem did not consist in the strange suggestion that computers do not have causal powers (mine certainly has) but in the question whether the computational processes *qua computational processes* have these powers (since their output is only the "0" in my example, not the position of a physical switch). In pancomputationalism, the stronger thesis about computing processes per se is claimed, and the weaker about running/actual/realized computing processes is defended.

A further example of this strategy is the defense of the stronger P1 via the weaker claim that the universe is processing information. It may well be true that information processing is an elementary feature of the universe, but information processing is information processing; computing is computing. Perhaps computing is one species of information processing among others (in some sense of 'information' it is), but why expand the one notion into the other? If we really want to say that all information processing is computational, is that a definitional remark or is this a discovery about information processing? If it is definitional, I might adopt my understanding of the thesis proposed here but I would then note that we now identify one unclear notion (computing) with another even less clear one (information processing); which does not look like a good strategy. In any case, all problems that beset the pancomputationalist approach also beset that of info-computationalism, plus the new ones associated with 'information'. If the remark is expressing a discovery, I would like to see the evidence for the claim that there is non-computational information processing does not exist. (In other words, I would come back to my first remark and wonder what a non-computational process would be, on the pancomputationalist account.)

Last but not least, the claim that the universe is computational looked quite strong when that term was understood as Turing computation, but then computation was dissolved into a much wider notion, the borders of which I cannot quite discern (I keep coming back to this issue). My worry about apparently non-computational processes in the world could not be countered because "everything is computing" is a priori, and we do not even want to take it as a reductionist claim ("everything is computing, deep down").

One example of my confusion is the interpretation of the remark by Feynman quoted above to support pancomputationalism. "In what sense of the word is that computing? In the sense of computation as a physical process, see Feynman's remark about physical computing from 2.2." This sounds circular and Feynman's remark does not help. I think it can well be read as *opposing* the idea that the universe computes (disregarding any context). He could be taken to say: Since it would take a computing machine infinite time, the "machinery" that is revealed in the end is not computational.

It truly is not clear to me how much can be explained with a wide notion of computing that somehow incorporates digital and analog, formal and physical, Turing and dynamic systems, etc. etc. I have nothing against these proposals, indeed my feeling is that some processes can usefully be described as computational (though even P6 in its generality is false) and many more as if they were. I also suspect that the metaphorical power of info-computationalism is strong enough to support an entire research program which will generate many interesting insights. Having said that, we have seen what happened to fruitful and successful research programs like classical AI or computationalism in the philosophy of mind that rested on weak foundations – they eventually hit the wall. I suspect that this will be the fate of info-computationalism also.

## 10.2.  GDC

First let us go back to Feynman. It is not a coincidence that this quote was used as the motto for the 2008 Midwest NKS Conference which gathered most prominent pancomputationalists. They interpret Feynman as saying that nature computes much more effective than any of our present machines. Moreover Feynman seems to imply that our going via mathematical models of physical phenomena might be the reason for that ineffectiveness.

Now the question of what is reasonable to understand as computation. For the nascent field of natural computation, we can apply the well known truth that our knowledge is in a constant state of evolution. Ray Kurzweil would even warn: Singularity is near, singularity where knowledge production exceeds our ability to learn (Kurzweil 2005). Moreover, by integrating/assimilating new pieces of knowledge, the whole existing knowledge structure changes. Atomism has changed substantially from the Democritus' original view. And yet it has not been refuted but only modified. Why? Because it was epistemically productive! Simply put, atomism has helped us to think, helped us to build new knowledge and to interact in different way with the physical world.

That is exactly what we expect from info-computationalism – to provide us with a good framework which will help our understanding of the world, including life and ourselves and our acting in that world. It is basically about learning and making sense of the world. Having history of several major paradigm shifts behind us, we have no reason to believe that info-computational framework is the absolutely perfect answer to all questions we may ask about the life, universe and everything but it seems to be the best research framework we have right now.

If pancomputationalism claims that the entire universe computes, a discovery of a process in the world which is impossible to understand as computational would falsify the pancomputational claim. Something changes, but we have no way to identify that process as computation. A stereotypical claim would be: writing a poem. That cannot possibly be a computational process! On which level of organization? I want to ask. On a level of neuroscience all that happens in the world while someone writes a poem is just a sequence of computational processes. Poets might find that level uninteresting, as well as they might find uninteresting the fact that the beautiful lady they sing of is made of atoms and void. But there are cases where we really want to know about how things work on a very basic level.

As a research program info-computationalism will either show to be productive or else it will die out. The only criterion for survival is how good it will be compared with other approaches. That development of a research programme is a slow but observable process. Following the number of articles, journals, conferences etc. dealing with unconventional computing, organic computing, or natural computing we can assess how active the field is. Subsequently we will also be able to follow its results.

Here is again a summary of what makes info-computationalist naturalism a promising research programme:

- Unlike mechanicism, info-computationalist naturalism has the ability to tackle as well fundamental physical structures as life phenomena within the same conceptual framework. The observer is an integral part of the info-computational universe.

- Integration of scientific understanding of the structures and processes of life with the rest of natural world will help to achieve "the unreasonable effectiveness of mathematics" (or computing in general) even for complex phenomena of biology that today lack mathematical effectiveness (Gelfand) – in sharp contrast to physics (Wigner).

- Info-computationalism (which presupposes pancomputationalism and paninformationalism) presents a unifying framework for common knowledge production in many up to know unrelated research fields. Present day narrow specialization into various isolated research fields has led to the alarming impoverishment of the common world view.

- Our existing computing devices are a subset of a set of possible physical computing machines, and Turing Machine model is a subset of envisaged more general natural computational models. Advancement of our computing methods beyond the Turing-Church paradigm will result in computing capable of handling complex phenomena such as living organisms and processes of life, social dynamics, communication and control of large interacting networks as addressed in *organic computing* and other kinds of *unconventional computing*.

- Understanding of the semantics of information as a part of the data-information-knowledge-wisdom sequence, in which more and more complex relational structures are created by computational processing of information. An evolutionary naturalist view of semantics of information in living organisms is given based on interaction/information exchange of an organism with its environment.

- Discrete and analogue are both needed in physics and so in physical computing which can help us to deeper understanding of their relationship.

- Relating phenomena of information and computation understood in interactive paradigm will enable investigations into logical pluralism of information produced as a result of interactive computation. Of special interest are open systems in communication with the environment and related logical pluralism including paraconsistent logic.

- Of all manifestations of life, mind seems to be information-theoretically and philosophically the most interesting one. Info-computationalist naturalism (pancomputationalism + paninformationalism) has a potential to support, by means of models and simulations, our effort in learning about mind and developing artifactual (artificial) intelligence in the direction of organic computing.

The spirit of the research programme is excellently summarized in the following:

> "In these times brimming with excitement, our task is nothing less than to discover a new, broader, notion of computation, and to understand the world around us in terms of information processing." (Rozenberg and Kari 2008)

**Acknowledgements**
The authors want to thank Ruth Hagengruber, Kaj Børge Hansen, Luciano Floridi and Mark Burgin for useful comments on earlier versions of this paper.